\documentstyle[psfig,conf_iap]{article}
\begin{document}
\heading{Are Damped Ly$\alpha$ Systems Rotating Disks ?}
\par\medskip\noindent
\author{C\'edric Ledoux$^{1}$, Patrick Petitjean$^{2}$, Jacqueline Bergeron$^{3}$}
\address{Observatoire Astronomique de Strasbourg, 11 Rue de l'Universit\'e, F--67000 Strasbourg, France}
\address{Institut d'Astrophysique de Paris, CNRS, 98bis Boulevard Arago, F--75014 Paris, France}
\address{European Southern Observatory, Karl-Schwarzschild-Stra$\ss$e 2, D--85748 Garching bei M\" unchen, Germany}
\begin{abstract}
We report on high spectral resolution observations of five damped Ly$\alpha$ 
systems whose line velocity profiles and abundances are analyzed. 
By combining these data with information from the literature, we study the 
kinematics of the low and high ionization phases of damped systems and 
discuss the possibility that part of the motions is due to rotation.
\end{abstract}
\section{The Si~{\sc ii}$\lambda$1808 and Fe~{\sc ii} profiles}
We study the general properties of the kinematics of the low and high 
ionization phases in a large sample of 26 damped Ly$\alpha$ systems 
(hereafter DLAS) covering the redshift range 1.17$-$4.38. 
It has been argued that high-redshift DLAS could represent a 
population of large, fast rotating disks mainly on the basis of the 
asymmetric shape of their low ionization species' absorption lines 
\cite{wol}\cite{pro}.
We note a strong correlation between the velocity broadenings of the 
Si~{\sc ii}$\lambda$1808 and Fe~{\sc ii}$\lambda$1608 transition lines for
apparent optical depths of 0.1 and 0.7 respectively (see Fig.~1, left panel). 
This shows that the physical conditions are quite homogeneous in the neutral
gas and that large variations of abundance ratios and 
thus large depletions onto dust grains in DLAS are unlikely.
Since the Fe~{\sc ii} transitions are easier to detect, we are able to 
compare velocities measured on {\it the same ion}.
%
\section{Velocity broadening and asymmetry}
%
We define an asymmetry parameter
$\eta$~=~$\left|V_1+V_2\right|/(V_1-V_2)$ that is close to zero for a
symmetric profile and close to one for a one-sided profile.
It can be seen on Fig.~1 (right panel) that the Fe~{\sc ii} line asymmetry is 
correlated with the velocity broadening up to
$\Delta V$~$\sim$~120~km~s$^{-1}$. For larger velocity broadenings however 
this correlation disappears. The systems with large $\Delta v$, such as the 
system at $z_{abs}$~=~3.151 toward Q~2233+131, are indeed very peculiar, 
with kinematics much more consistent with random motions than ordered disks. 
They show sub-systems as those expected if the objects are in the process of 
merging. Therefore the claim that CDM models should be rejected has to be 
considered with caution. Indeed using hydro-simulations, Haehnelt et al.
(1997)
\cite{hae} have reached the same conclusion. Although more work has to be 
done especially investigating comparisons with simple kinematical models, 
the overall picture seems to be more in favor of the DLAS to be aggregates of
dense knots with complex kinematics rather than ordered disks. 
This does not rule out however the 
possibility that part of the kinematics could be due to rotation.
The correlation shown in Fig.~1 (right panel) is suggestive of
rotational motions in sub-systems on scales smaller 
than $\Delta V$~$<$~120~km~s$^{-1}$. 
The kinematics of the low and high ionization species themselves is found to 
be statistically correlated though the high ionization phase has a much more 
disturbed kinematical field than low ions. This should be taken into account 
in any model of high-redshift damped Ly$\alpha$ systems.

\begin{figure}[t]
\centerline{\hbox{
\psfig{figure=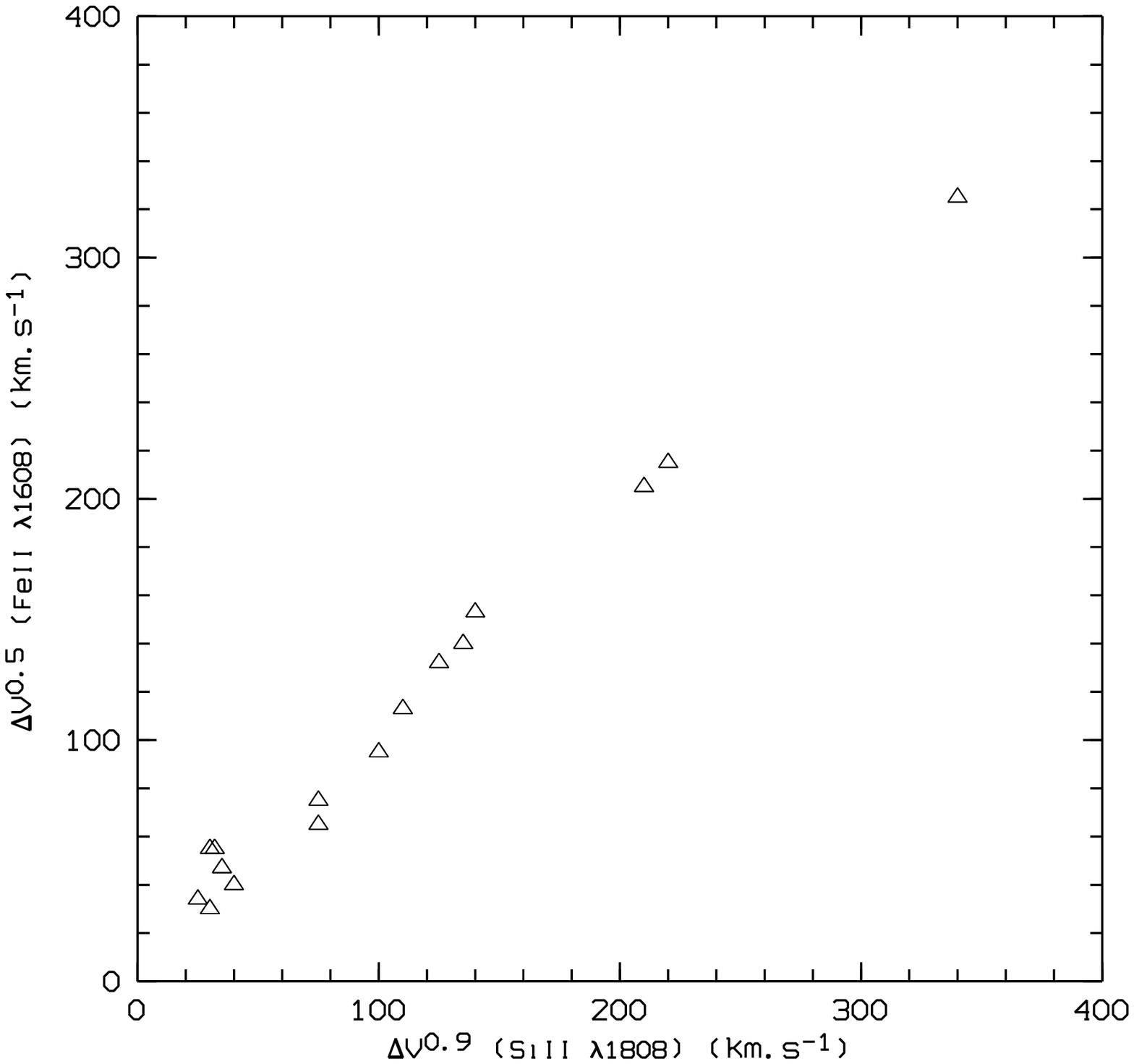,height=6.cm}
\psfig{figure=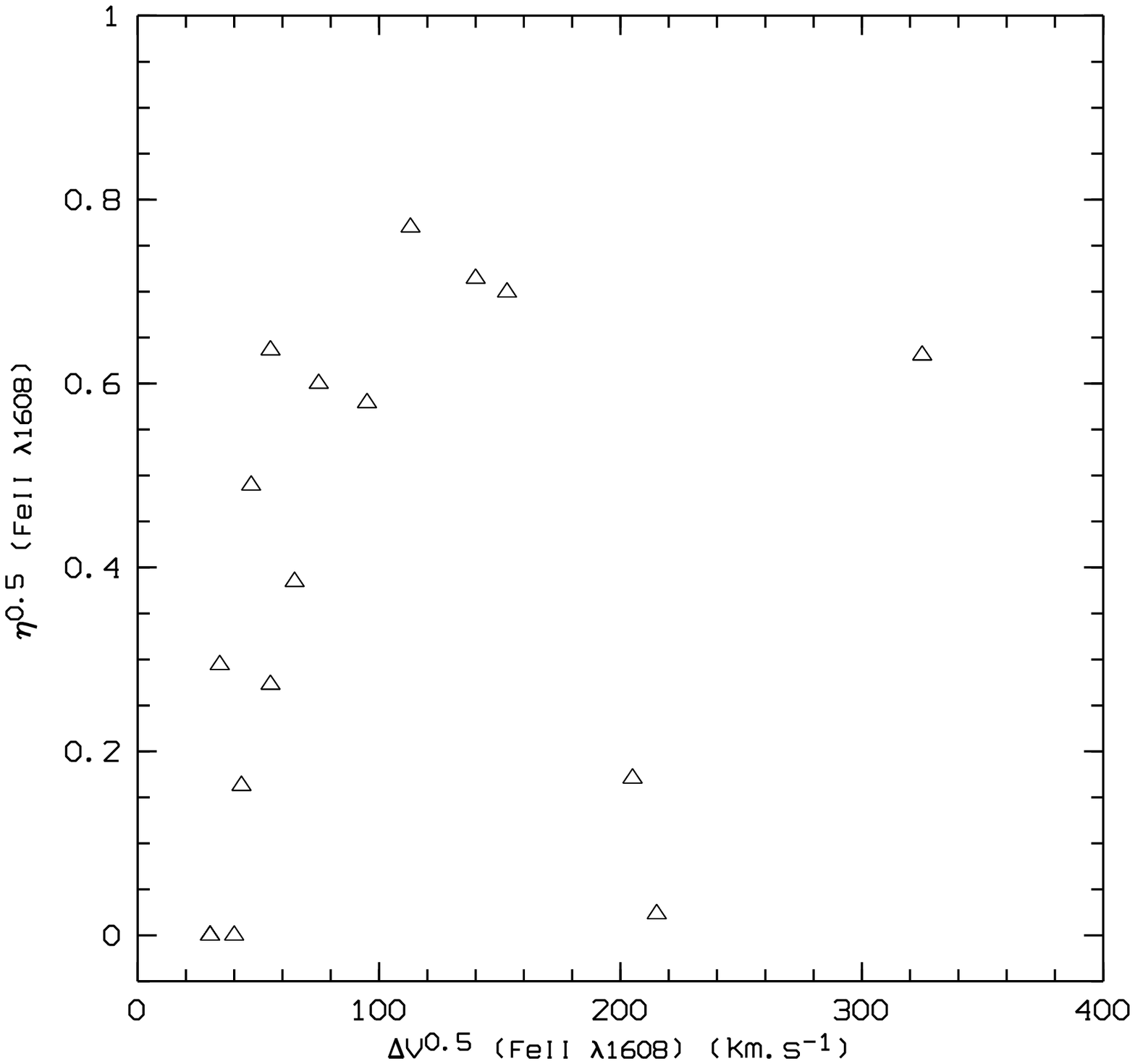,height=6.cm}}}
\caption[]{Velocity broadening of the Fe~{\sc ii}$\lambda$1608 lines at
$\tau_{\nu}$~$\sim$~0.7 versus that of Si~{\sc ii}$\lambda$1808 at
$\tau_{\nu}$~$\sim$~0.1 (left panel) and line asymmetry versus the velocity
broadening for Fe~{\sc ii}$\lambda$1608 measured at $\tau_{\nu}$~$\sim$~0.7
(right panel).}
\end{figure}
\acknowledgements{The participation of C. Ledoux to the conference has been
funded by the E.C.}
\vfill
\end{document}